\documentclass[fleqn,usenatbib]{mnras}   
\usepackage{graphicx,times,epsf}            
\usepackage{longtable}
\usepackage{natbib}
\usepackage{amssymb}
\usepackage{amsmath}
\usepackage{ulem}
\usepackage{soul,xcolor}

\begin{document}

\title{Transit Timing Variation of K2-237b: Hints Toward Planet Disk Migration}

\author[Yang et al.]{Fan Yang\href{https://orcid.org/0000-0002-6039-8212}{\includegraphics[scale=0.1]{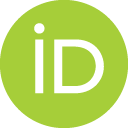}}$^{1,}$$^{2,}$$^{8}$\footnotemark[1], Richard J. Long\href{https://orcid.org/0000-0002-8559-0067}{\includegraphics[scale=0.1]{ORCIDiD_icon128x128.png}}$^{3}$\footnotemark[2], Eamonn Kerins\href{https://orcid.org/0000-0002-1743-4468}{\includegraphics[scale=0.1]{ORCIDiD_icon128x128.png}}$^{4}$,
 Supachai Awiphan\href{https://orcid.org/0000-0003-3251-3583}{\includegraphics[scale=0.1]{ORCIDiD_icon128x128.png}}$^{5}$,\newauthor
Su-Su Shan\href{https://orcid.org/0000-0002-5744-2016}{\includegraphics[scale=0.1]{ORCIDiD_icon128x128.png}}$^{2,}$$^{1}$, Bo Zhang\href{https://orcid.org/0000-0002-6434-7201}{\includegraphics[scale=0.1]{ORCIDiD_icon128x128.png}}$^{1}$, Yogesh C. Joshi 
\href{https://orcid.org/0000-0001-8657-1573}{\includegraphics[scale=0.1]{ORCIDiD_icon128x128.png}}$^{6}$, Napaporn A-thano\href{ https://orcid.org/0000-0001-7234-7167}
{\includegraphics[scale=0.1]{ORCIDiD_icon128x128.png}}$^{5}$, 
\newauthor
Ing-Guey Jiang
\href{https://orcid.org/0000-0001-7359-3300}{\includegraphics[scale=0.1]{ORCIDiD_icon128x128.png}}${^7}$,
Akshay Priyadarshi\href{https://orcid.org/0000-0003-1143-0877}{\includegraphics[scale=0.1]{ORCIDiD_icon128x128.png}}$^{4}$, 
and Ji-Feng Liu\href{https://orcid.org/0000-0002-2874-2706}{\includegraphics[scale=0.1]{ORCIDiD_icon128x128.png}}$^{1,}$$^{2}$\\
$^{1}$National Astronomical Observatories, Chinese Academy of Sciences, 20A
     Datun Road, Chaoyang District, Beijing 100101, China\\
$^{2}$School of Astronomy and Space Science, University of Chinese Academy of Sciences,
Beijing 100049, China\\
$^{3}$Department of Astronomy, Tsinghua University, Beijing 100084, China\\
$^{4}$Jodrell Bank Centre for Astrophysics, Department of Physics and Astronomy, The University of Manchester, Oxford Road, Manchester M13 9PL, UK\\
$^{5}$National Astronomical Research Institute of Thailand, 260 Moo 4, Donkaew, Mae Rim, Chiang Mai 50180, Thailand\\
$^{6}$Aryabhatta Research Institute of Observational Sciences (ARIES), Manora Peak, Nainital 263001, India\\
$^{7}$Department of Physics and Institute of Astronomy, National Tsing-Hua University, Hsinchu 30013, Taiwan\\
$^{8}$D\'epartement d'Astrophysique/AIM, CEA/IRFU, CNRS/INSU, Univ. Paris-Saclay, Univ. de Paris, 91191 Gif-sur-Yvette, France\\
}

\maketitle

\begin{abstract}
Hot Jupiters should initially form at considerable distances from host stars and subsequently migrate towards inner regions, supported directly by transit timing variation (TTV). 
We report the TTV of K2-237b, using reproduced timings fitted from \textit{Kepler} K2 and \textit{TESS} data. The timings span from 2016 to 2021, leading to an observational baseline of 5 years. The timing evolution presents a significant bias to a constant period scenario. The model evidence is evaluated utilizing the Bayesian Information Criterion (BIC), which favours the scenario of period decay with a $\Delta$BIC of 14.1. The detected TTV induces a period decay rate ($\dot{P}$) of -1.14$\pm$0.28$\times$10$^{-8}$ days per day ($-$0.36 s/year). Fitting the spectral energy distribution, we find infrared excess at the significance level of 1.5 $\sigma$ for WISE W1 and W2 bands, and 2 $\sigma$ level for W3 and W4 bands. This potentially reveals the existence of a stellar disk, consisting of hot dust at 800$\pm$300 K, showing a $L_{dust}/L_{\ast}$ of 5$\pm$3$\times$10$^{-3}$. We obtain a stellar age of 1.0$^{+1.4}_{-0.7}$$\times$10$^{9}$ yr from isochrone fitting. The properties of K2-237b potentially serve as a direct observational support to the planet disk migration though more observation are needed.
\end{abstract}
\footnotetext[1]{Fan Yang: Fan.YANG@cea.fr; sailoryf1222@gmail.com}
\footnotetext[2]{Richard J. Long: rjlastro@yahoo.com}

\section{Introduction}

Migration is believed to explain the presence of the giant planets orbiting close to their host stars, commonly known as hot Jupiter \citep{Lin1996, Dawson2018}. Transit timing variations play a crucial role in supporting this migration theory, yet their detection requires high-precision light curves and a long timing baseline \citep{2017wasp12b, 2021wasp12bTESS}. The {\it Kepler} mission \citep{Kepler2010} and subsequent K2 mission \citep{K2}, and Transiting Exoplanet Survey Satellite \citep[TESS;][]{Ricker2015}, have provided invaluable data for TTV research, leading to numerous intriguing discoveries \citep{Baluev2020, Wang2024, TransitFit}.

It has the potential to distinguish and understand different tunnels of giant planet evolution, by jointly analyzing the TTV evidence and other properties of star-planet system \citep{Yangwasp161}. For example, a hot Jupiter at a low eccentricity ($e$) orbit, embedded in a young stellar system, supports the disk migration model, which would obtain extra significance if with direct disk evidence \citep{Papaloizou2000, Nelson2000}. In another scenario, the hot Jupiter having an old host star and a large $e$ orbit would indicate a tidal migration path, i.e., decaying due to the tidal forces after a phase of dynamical scattering \citep{Rasio1996, Mardling2007, Lendl2014}. Hot Jupiters, e.g., WASP-12b, WASP-161b, are reported with tidal migration evidence \citep{2011wasp12b, Yangwasp161, Bai2022,Leonardi2024}, whereas TTV detection relating to disk migration is still absent.

K2-237b is a giant planet orbiting an F-V type star with an orbital period of 2.18 days, which was discovered through K2 light curves \citep{Soto2018, Smith2019}. Notably, the transit ephemeris of this planet has undergone frequent refinements \citep{Ikwut-Ukwa2020,shan2023, Ivshina2022, Patel2022}, including significant differences in periods among studies conducted by the same research team \citep{Edwards2021, Kokori2022, Kokori2023}. This raises concerns regarding the consistency of the observational timing measurements or the accuracy of the timing evolution model assuming a constant period.

In this work, we report the detection of the transit timing variation of K2-237b, and clues of the existence of a stellar disk. The organization of the paper is structured as follows. In Section 2, we described the data reduction and timing acquirement. Section 3 presents the timing evolution and possible physical origins. Section 4, briefly summarizes the results. 

\section{Data Reduction and Timing Measurements}

\subsection{Photometric Light Curves}

\begin{figure}
    \includegraphics[width=3.5in]{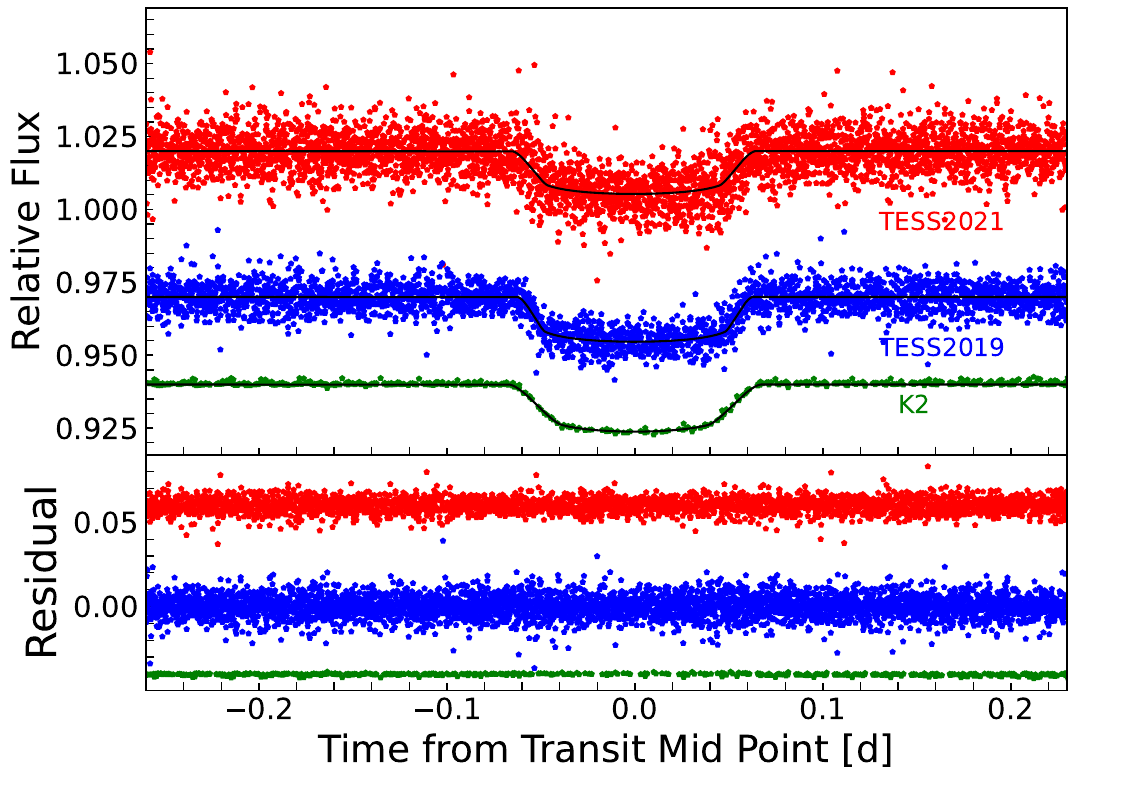}
  \caption{Folded light curves of K2-237b, along with the residuals shown in the bottom panel. The light curves are colour-coded with red, blue, and green points, representing the TESS observations in 2021, 2019, and K2 observations, respectively, arranged from top to bottom. The fitted transit models are present in black lines. For clarity, the light curves have been vertically shifted.
  }
\label{image: lc} 
\end{figure}

We use light curves generated by sophisticated data reduction pipeline \citep[Presearch Data Conditioning, PDC; ][]{PDC} for both TESS and K2 data, for purposes of convenience in timing comparison among different works (as shown in Figure \ref{image: lc}). It is notable that, obtaining pixel-by-pixel images and generating the light curve ourselves would not induce significant differences in the transit timings, according to our previous experiences \citep{Yangwasp161,yangxo3b,Yang2022RAA,shan2023}.

TESS boasts a resolution of 21 arcseconds per pixel and occupies four cameras with a total field of view (FOV) of 24×96 deg$^{2}$ \citep{Ricker2015}. TESS continuously observes a scheduled sky area every 27 days, referred to as a sector. TESS light curves are available with different cadences, i.e., 30 minutes and 2 minutes for the 2019 observation; and 2 minutes and 20 seconds for the 2021 observation. We utilize the cadence of 2 minutes, considering the consistency among observations in different years. We previously reported a cadence of 2 minutes would not induce any parameter uncertainties when modelling TESS light curves \citep{yangLD, Yangatmos}.

K2 possesses an FOV of $\sim$ 100 deg$^{2}$ with a pixel resolution of 4 arcseconds \citep{K2}. K2 maintains a constant boresight throughout each campaign, i.e., 80 days. K2 light curve (in 2016) has a cadence of 30 minutes which could induce extra uncertainties to some transit parameters, e.g., transit depth \citep{Kipping2010, Yangatmos}, whereas the timings should not be significantly influenced due to the symmetry of the light curve welly enclosing the ingress and egress.

\subsection{Modelling the Transit Light Curves}

A detrending is required for light curve analysis, to remove the trends relating to instrumental effects and possible stellar activities. We apply two different approaches in detrending, i.e., one applying Gaussian Process \citep[GP;][]{celerite} and the other applying a simple linear fit \citep{GPdetrending, Yangatmos}. Before either detrending, we perform an outlier clipping, i.e., removing the points beyond five times the mean absolute deviation (MAD) of local 11 points for TESS data and local 5 points for K2 data.

The light curve in each epoch is performed with detrending and transit modelling simultaneously which has been described in previous publications \citep{yangLD, Yangatmos, yangxo3b}. We briefly describe the modelling process here and elaborate on GP detrending.

We apply a classic transit model \citep{Mandel_Agol2002} and depict light curve trends by an extra scaling factor. The transit fit takes a circular parameter set, being consistent with the discovery papers \citep{Soto2018, Smith2019}. The free parameters enclose the transit mid-point ($T0$), the radius ratio of the planet to the host star ($R_{p}/R_{\ast}$), the semi-major axis in unit of stellar radius ( $a/R_{\ast}$), and the limb darkening (LD) parameters.

The priors (except LD) for the transit parameters refer to the previous publication and utilize Gaussian format \citep{Soto2018}. The centres take the reported median value, yet the uncertainties are set as 10 times the reported 1 $\sigma$ region. The LD prior takes the centre values of (0.38, 0.26) for K2 band and (0.28, 0.24) for TESS band, predicted by \citep{exoctk}, based on ATLAS stellar model \citep{ATLAS}. The uncertainties are set as 0.05, according to the previous experience \citep{yangLD}.

\begin{table}
\setlength{\tabcolsep}{0.5mm}
\begin{center}
\caption{K2-237b transit timings with 1 $\sigma$ significance region.}
\label{table: timings}
\begin{tabular}{cccc}
  \hline
 \hline
\multicolumn{4}{c} {Transit Mid-points (BJD-2457684.8101)} \\
\hline 
\multicolumn{4}{c} {Timings Involved in Modeling}\\
K2                &     TESS 2019          &   TESS 2021           & El Sauce \\
\hline 
\multicolumn{4}{c} {GP Detrending Timings} \\
-19.6252$\pm$0.0002 &  950.7145$\pm$0.0004 &  1696.4556$\pm$0.0004 & \\
\hline 
\multicolumn{4}{c} {NGP Detrending Timings} \\
-6.5411$\pm$0.0003 &  959.4367$\pm$0.0005 &  1696.4550$\pm$0.0005 &   \\
\hline 
\multicolumn{4}{c} {Archival Timings} \\
 0.0000$\pm$0.0001 &  955.0750$\pm$0.0004 &  1679.0113$\pm$0.0007 & 904.9237$\pm$0.0004   \\
\hline 
\hline
\multicolumn{4}{c} {Timings not Involved in Modeling}\\
\multicolumn{4}{c} {GP Detrending Timing of Individual Epochs} \\
-23.9864$\pm$0.0015 &  -21.8110$\pm$0.0020 &  -6.5349$\pm$0.0018 &  4.3668$\pm$0.0012  \\
10.8918$\pm$0.0018 &  15.2665$\pm$0.0018 &  19.6266$\pm$0.0014 &  39.2433$\pm$0.0013 \\
946.3535$\pm$0.0010 &  948.5342$\pm$0.0010 &  950.7140$\pm$0.0010 &  952.8946$\pm$0.0011 \\
 959.4365$\pm$0.0010 &  961.6169$\pm$0.0011 &  963.7984$\pm$0.0011 &  965.9787$\pm$0.0010 \\
 1679.0100$\pm$0.0018 &  1681.1901$\pm$0.0015 &  1683.3609$\pm$0.0019 &  1685.5545$\pm$0.0015 \\
 1687.7296$\pm$0.0016 &  1696.4590$\pm$0.0014 &  1698.6414$\pm$0.0017 &  1700.8056$\pm$0.0020 \\
\hline
\end{tabular}
\end{center}
\begin{flushleft}
\end{flushleft}
\end{table}

The GP detrending applies a SHOTerm kernel, incorporating three hyper-parameters \citep{celerite, Maxted2022}. 
The prior of the GP hyper-parameters are obtained by an iteration method, referring to the approach of empirical Bayesian \citep{EmpiricalBayes,yangLD, Maxted2022}. Initially, a fit is conducted to the light curve without any priors. Subsequently, the fitted result serves as the prior for the final fit, with the uncertainty expanded twofold to enhance model flexibility and prevent overfitting.

The linear detrending, referred to as NGP hereinafter, obscures the light curves proximal to the transit by a duration of 0.6 days. It characterizes the trends through a linear function, as described in \citep{yangLD, Yangatmos}. Tests conducted with higher-order polynomial functions, up to the fifth order, reveal a negligible difference.

\citet{Soto2018} report a K2-237 rotation period of 5.11$\pm$0.02 days, implying surface features, e.g., stellar spots. We find that any potential fluctuations in the light curve arising from stellar structures, if they exist, are minimal. Consequently, any reshaping of the light curve is negligible. Furthermore, any residual shaping of the light curve that may persist is effectively mitigated through GP detrending, which underscores one of the key advantages of employing this technique. A comparative assessment between light curves processed with and without GP detrending fails to yield any notable discrepancies. The timing measurements acquired both with and without GP detrending align well with their respective uncertainty margins.

\section{TTV Evidence and Interpretation}

\subsection{Timing Evolution}

A Monte-Carlo Markov Chain (MCMC) technique is applied when fitting the light curves \citep{pymc, emcee}. This MCMC approach involved an initial burn-in phase of 50,000 steps, followed by 80,000 subsequent steps for sampling. We exclude light curves lacking ingress or egress measurements, particularly in the case of K2 light curves due to their 30-minute sampling rate, which inadequately constrains the transit timings. Additionally, we discard light curves exhibiting apparent abnormal photometric measurements, as they introduce significant uncertainties in the transit modelling process. The timings derived from Gaussian Process detrending in each epoch are listed in Table \ref{table: timings}, revealing notable fluctuations attributed to systematic effects. We note that our timings in each epoch are consistent with the results from other work \citep{Edwards2021, Ivshina2022}. In addition, the timing accuracy of our pipeline has been tested and proved by our previous publications \citep{Yangwasp161,yangxo3b,Yang2022RAA,shan2023}.

For timing evolution analysis, we used the timings obtained from folded detrended light curves (as shown in Figure \ref{image: lc}), in order to reduce the systematic involved in the single epoch result (as shown in Table \ref{table: timings}). The folding period comes from \citet{shan2023}, referring to a refined constant period obtained from K2, TESS light curves, which is consistent with recent publications at the level of 0.000002 days \citep{Ikwut-Ukwa2020, Ivshina2022, Kokori2023}. According to further tests, the timing drift due to different folding periods relating to different timing evolution models is $\sim$ 0.00003 days within one TESS sector or K2 campaign, which would not induce a significant difference in folding the light curves. The modelling of the folded light curve is the same as the process of single epoch light curves but without detrending factors.

\begin{figure}
\centering
    \includegraphics[width=3.5in]{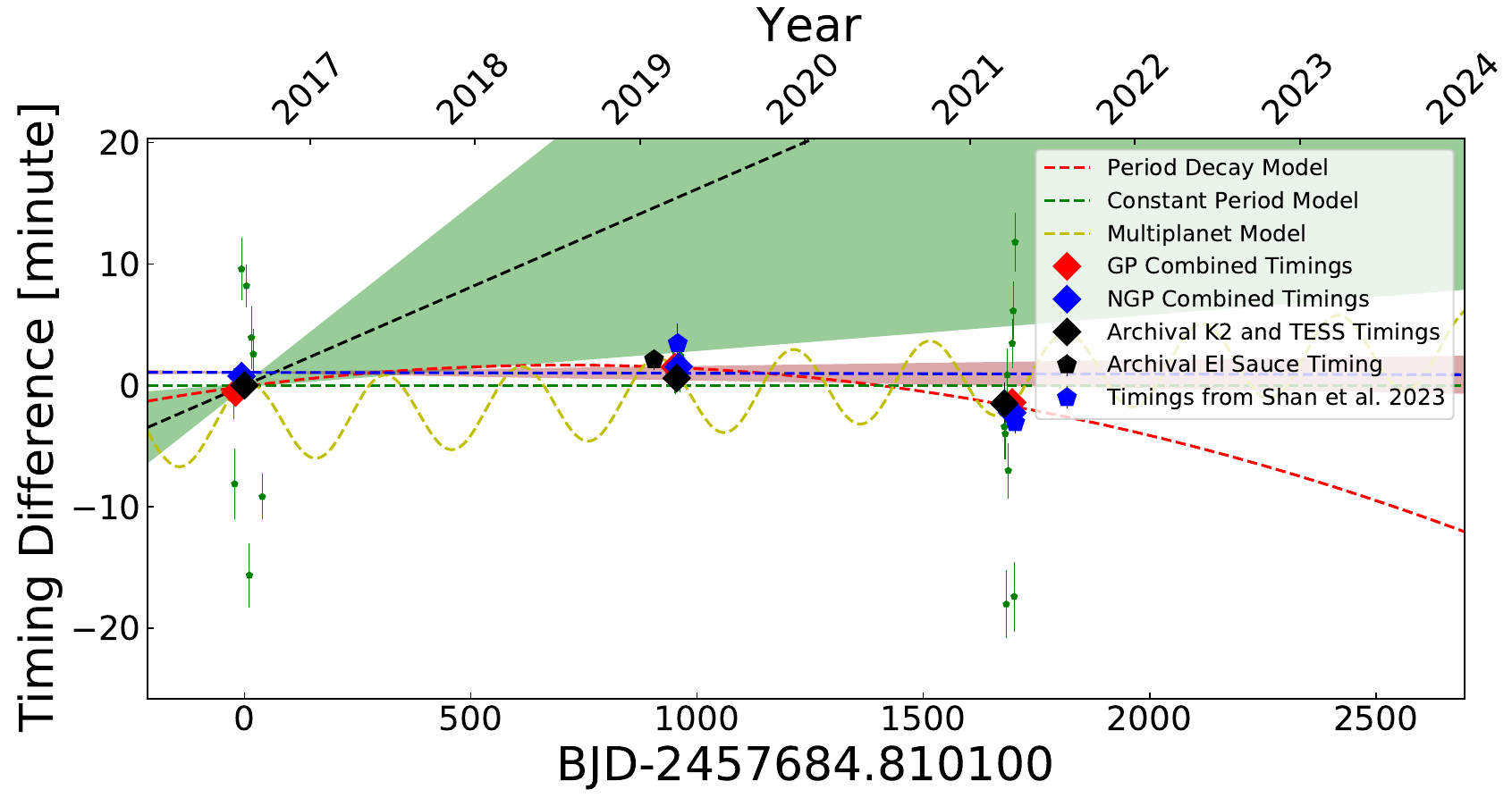}
  \caption{Timing evolution of K2-237b transit. The black line and green region present the ephemeris with 1 $\sigma$ region, reported by \citet{Soto2018}. The blue line and magenta region show a refined ephemeris from \citet{Ikwut-Ukwa2020}. Diamonds in different colours indicate K2 and TESS timings; red, timings from combined light curves (K2, TESS 2019, TESS 2021 in time series) with GP detrending; blue, timings from combined light curves without GP detrending; black, archival timings. The black pentagon depicts archival timing from El Sauce \citep{Edwards2021}. Timings shown by diamonds and black pentagon are evolved in timing analysis. The red line represents the period decay model, the green line signifies the constant period model, and the yellow line illustrates the multi-planet interaction model. Additionally, the blue pentagons depict the TESS timings reported in our prior study \citep{shan2023}. Notably, the inclusion of these timings from \citet{shan2023} would enhance the preference significance for the period decay model. However, they are excluded from our timing evolution analysis due to their dependency on non-GP detrending timings obtained in this work. For comparison, we also present the transit timings of each epoch using small green pentagons.}
\label{image: timingevolution} 
\end{figure}

During the timing evolution analysis, the timings derived from both Gaussian Process (GP) and Non-Gaussian Process (NGP) detrending are treated as equally valid. The observed differences are insignificant, i.e., within 1 $\sigma$ (as demonstrated in Figure \ref{image: timingevolution}).
Moreover, we incorporate previously reported timings that were derived from a single TESS sector or K2 campaign, allowing a comprehensive consideration of various timing modelling pipelines \citep{Soto2018, TOIcatalog, Patel2022}.  
These timings are applied with a scaling factor, ensuring the timings from the same observation would not be over-weighted in the timing evolution analysis. The scaling factor follows the error propagation law, i.e., $\sigma_{scale}$=$\sigma_{origin}\times\sqrt{N}$, where 
$\sigma_{origin}$ is the unscaled uncertainty obtained from transit modelling, $N$ denotes the number of timings from different data reduction methods, and $\sigma_{scale}$ is the scaled uncertainty used in the timing evolution analysis. Timing measurements range from 2016 to 2021, forming a baseline of 5 years in total.

Modelling the timing measurements as described above and archival timing from El Sauce \citep{Edwards2021}, we find the evolution significantly deviating from a linear function, referring to a constant period model (as shown in Figure \ref{image: timingevolution}). we enclose models of a quadratic function, referring to period decaying, and a sinusoidal function, referring to multi-planet interacting, which are discussed in the next section. We occupy the Bayesian Information Criterion \citep[BIC, details in][]{BICmostcited} in the evaluation of the model evidence. The BIC for the constant period model is 25.2, the period decay model is 11.1, and the multi-planet interaction model is 18.7, indicating a significant preference towards the period decay model. 
In addition, we fit the archival timings obtained from one TESS sector, K2 campaign and El Sauce, in total 4 measurements, using the same functions. The preference for the period decay model does not change. 

\subsection{Possible Physical Origins}

The planet-disk interactions function as torques to exchange angular momentum between disk and planet orbit, propelling the planet towards its host star \citep{Lin1996, Dawson2018}. This mechanism should be the most plausible explanation for the observed TTV which can be the first direct observational evidence of a migrating hot Jupiter due to disk interactions. The period derivative derived from fitting the timing measurements is -1.14$\pm$0.28$\times$10$^{-8}$ days per day ($-$0.36 s/year).
The equivalent dissipation timescale of 0.5 Myr is significantly shorter than what would be expected for the competing period decay model that is motivated by tidal effects \citep{Goldreich1966, Hut1981, Mahmud2023}. The efficiency of tidal migration can be boosted by mechanisms such as magnetic interaction and diffuse tides \citep{Strugarek2015, Yu2022, Wei2024}, these factors may help explain the rapid period decay observed. After balancing the evidence, we choose to favour disk migration, however this does not rule out tidal migration completely.

\begin{figure}
\centering
    \includegraphics[width=3.5in]{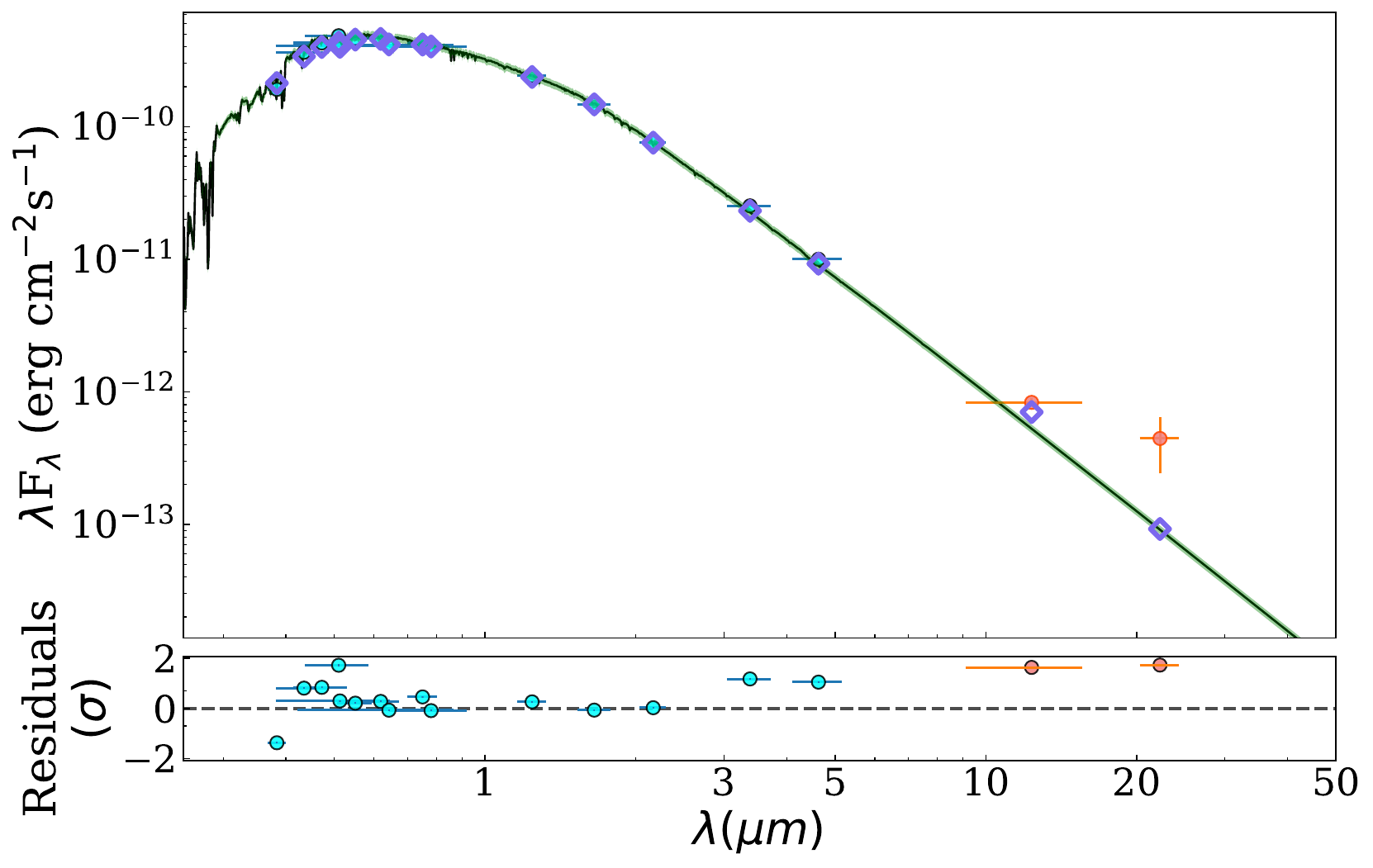}
    \includegraphics[width=3.5in]{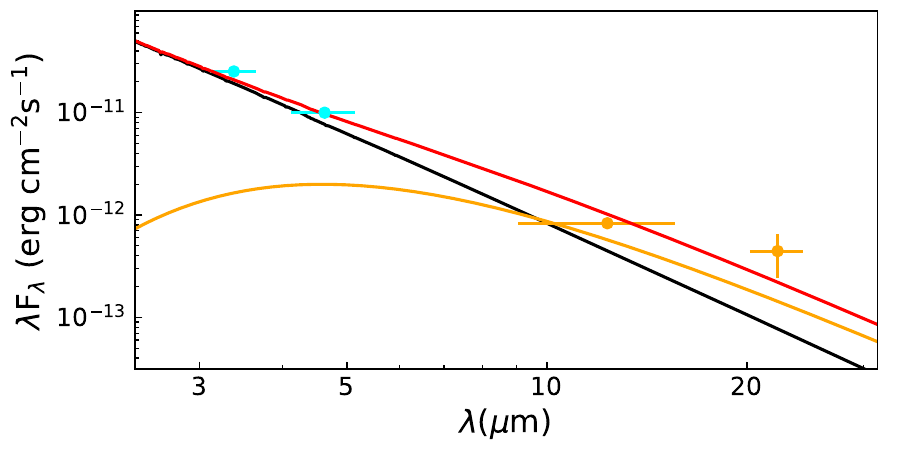}
  \caption{The spectral energy distribution of K2-237. Top: The synthetic flux is generated from the ATLAS stellar model and indicated by purple diamonds. The ATLAS stellar model is depicted by the solid black line. The green area encompasses the 1$\sigma$ uncertainty region, dominated by the uncertainty of the synthetic model.  
The observational fluxes are represented by cyan circles. WISE W3 and W4 band fluxes are highlighted in orange for analyzing infrared excess.
In the middle panel, the residuals are present, normalized to the total uncertainty in each band. 
Bottom: The disk-involved model with the x-axis zoomed in on WISE bands. The orange line represents the thermal disk emission model, while the red lines present the composite model that incorporates both stellar and disk emissions.
  }
\label{image: SED} 
\end{figure}

For disk migration, the eccentricity ($e$) should damp before the giant planet becomes hot and maintains law \citep{Cresswell2007, Duffell2015}, which just corresponds to the observed $e$ of K2-237b. 
Interestingly, our analysis reveals indications of a stellar disk, inferred from the infrared excess observed in the WISE bands \citep{Wright2010}. We detect a significance of 1 $\sigma$ at W1 and W2, and 2 $\sigma$ at W3 and W4, as depicted in Figure \ref{image: SED}. The spectral energy distribution (SED) is generated utilizing ARIADNE \citep{Vines2022}, which fits to the archival observational photometric measurements \citep{Fitzpatrick1999, Tycho2, SDSS, 2mass, Wright2010, Gaiadr3}, leveraging ATLAS stellar model \citep{ATLAS}.

The excess in WISE flux can be fitted by incorporating a straightforward thermal disk component \citep{Lawler2012, Mann2016}.
We fit the WISE flux using the MCMC method, revealing the presence of hot dust with a temperature of 800$\pm$300 K and a luminosity ratio of dust to the star ($L_{dust}/L_{\ast}$) amounting to 5$\pm$3$\times$10$^{-3}$ (as illustrated in Figure \ref{image: SED}). These disk parameters align with the empirical ranges established in prior literature \citep{Meyer2008, Hughes2018}. 
The SED model that incorporates a stellar disk exhibits a notably smaller BIC value of 20.3, compared to the non-disk SED model. A scaling factor has been applied to the model error, making the reduced $\chi^{2}$ for disk-involved to be 1 (in WISE bands). The reduced $\chi^{2}$ of the non-disk model is 3.55.

While estimating the stellar age remains challenging, we have discovered compelling evidence that K2-237 is a young star, aligning with indications of the existence of a stellar disk. Employing the isochrone table from \citep{Dotter2016}, we derived a stellar age of 1.0$^{+1.4}_{-0.71}$$\times$10$^{9}$ yr, based on the stellar parameters obtained through SED fitting \citep{Dotter2016}. This finding is in agreement with the results reported by \citep{Ikwut-Ukwa2020}, who also employed the isochrone method. Furthermore, the reported stellar rotation period of 5.11$\pm$0.02 days suggests a young stellar age, as indicated by \citep{Soto2018}. Quantitatively, \citet{Smith2019} derived a stellar age of 1.2$\pm$0.7 Gyr using the gyrochronology relation from \citet{Barnes2010}. The debris disk should still be detectable at infrared bands when having a stellar age of a few hundred Myr \citep{Meyer2008, Moor2011}. 
It is important to note that both isochrone fitting and gyrochronology methods exhibit limited sensitivity and precision when applied to young stars, thereby rendering their age determinations somewhat uncertain. For instance, K2-237, being an F6-type star, lies near the threshold where stellar magnetic activity significantly slows down rotation. Consequently, its interpretation varies across different studies, with some presenting it as representative of the average rotation period for an F6 star \citep{Nielsen2013}. We refrained from adjusting the age uncertainty ranges as it is beyond the scope of this work.

Intriguingly, the estimated disk temperature of 800$\pm$300 K surpasses that of typical debris disks and aligns with protoplanetary disks \citep{Chen2006, Krivov2011, Williams2011, Chen2014}. If the disk is a debris disk, it should be located in closer proximity to its host star. This proximity implies a potential orbital alignment with the observed hot Jupiter, suggesting a dynamic connection between the two. The coupling between the disk and the hot Jupiter could be indicative of disk formation through the stripping of the latter's materials due to tidal interactions. Nevertheless, the gas content in such disks typically contributes minimally to the continuum flux, necessitating an implausibly large amount of stripped material from the hot Jupiter's atmosphere under standard assumptions. Alternatively, the disk may be a continuation of the early protoplanetary disk phase, with its longevity enhanced by the presence of the hot Jupiter, as posited in \citep{Baruteau2014, Michel2021}. 
The current observations fall short in the determination of the precise gas fraction within the observed disk, indicating the need for additional observations, e.g., the CO emission line.

The pivotal determinant of whether migration can be initiated by the disk lies in its mass. The planet-disk interaction plays a crucial role in transferring the orbital energy of the planet to the disk, a process that can occur via mechanisms such as type II migration in the context of protoplanetary disks \citep{Baruteau2014, Williams2011} or particle scattering in debris disks \citep{Friebe2022}. Notably, the observed disk exhibits a higher luminosity fraction than previously reported debris disks hosting transiting exoplanets \citep{Krivov2011}, potentially signalling a more substantial stellar disk.
Furthermore, \citet{Friebe2022} conducted simulations specifically focusing on the migration of Jupiter-like planets within massive debris disks, offering insights into the dynamics at play in such systems.

Multi-planet interaction is a potential alternative origin of the detected timing evolution for K2-237b (as shown in Figure \ref{image: timingevolution}), though the $\Delta$BIC presents a `strong' preference to the period decay model \citep{BIC}. The fitted sinusoidal function gives a cycling period of 174.5$\pm$3 days and an amplitude of 1.7$\pm$0.5 minutes (Figure \ref{image: timingevolution}). A small planet hidden at an inner orbit near 2:1 mean motion resonance can produce the signal as detected. According to the relation from \citet{Lithwick2012}, the small planet should have a mass range of 0.5 to 2 earth mass.

The Doppler effect caused by a stellar companion should not be the origin of the observed period change. \citet{Smith2019} fitted K2-237 radial velocity (RV) residual with the effect of giant planet removed, using a linear function. The residual does not show any significant trend, indicating no radial acceleration evidence.

The observed TTV of K2-237b should not be attributed to apsidal precession, despite its discussion as a possible explanation for detected TTVs in other systems \citep{2017wasp12b, yangxo3b, Yangwasp161}. A detectable precession requires a high $e$ which is not the case of K2-237b \citep{Ragozzine2009, Soto2018, Smith2019, Ikwut-Ukwa2020}.

\subsection{The Requirement and Schedule for Further Observation}

More observations are imperative to discriminate among scenarios of disk migration and multiplanet interaction. The difference in transit timings should be more than 10 minutes since the middle of 2023, as shown in Figure \ref{image: timingevolution} (BJD$\sim$2500). 

We have scheduled more transit measurements utilizing the Spectroscopy and Photometry of Exoplanet Atmospheres Research Network \citep[SPEARNET][]{morgan19, A-thano2022, TransitFit}, in particular, the equipment of 0.7-m Thai Robotic Telescope at the Gao Mei Gu Observatory (TRT-GAO), China; the Spring Brook Observatory (TRT-SBO), Australia; and the Cerro Tololo Inter-American Observatory (TRT-CTO), Chile. 
The FOVs are 21 $\times$ 21 arcmin$^{2}$ for TRT-GAO, 
28 $\times$ 28 arcmin$^{2}$ for TRT-SBO, and 10 $\times$ 10 arcmin$^{2}$ for TRT-CTO.
The observations are conducted using the photometric filter of \textit{R}-band. We apply an observation strategy requiring the coverage of at least ingress or egress for the transit to improve the timing precision. Light curves are already obtained for two transits, in September 2023 and March 2024. The timing precision would reach 1 minute when combining transit light curves in $\sim$ 5 epochs, thereby, leading to distinguishing significance at higher than 10$\sigma$ (in Figure \ref{image: timingevolution}). 

\section{Summary}

We report the detection of significant transit timing variation of K2-237b, using timings obtained from K2 and TESS light curves, and archival El Sauce timing \citep{K2, Ricker2015, Edwards2021}. The observation spans from 2016 to 2021, forming a timing baseline of 5 years. We obtained the K2 and TESS timings, utilizing the classic transit model, detrending methods of both the Gaussian Process and linear function, and MCMC techniques \citep{Mandel_Agol2002, celerite}. The timing evolution is best explained by a period decay scenario with the least BIC of 11.1, compared to the constant period BIC of 25.2.
Fitting to the timing measurements, we derive a period derivative of -1.14$\pm$0.28$\times$10$^{-8}$ days per day, equivalent to $-$0.36 s/year.

The disk-planet interaction stands as our supported physical explanation for the observed Transit Timing Variations (TTVs), with tidal migration continuing to be a viable alternative hypothesis.
We present tentative evidence of a stellar disk, by fitting the SED of K2-237 \citep{Vines2022}. The flux exhibits an excess in the WISE bands, reaching significance levels of 1.5 $\sigma$ for W1 and W2, and 2 $\sigma$ for W3 and W4. This excess can be interpreted with a thermal hot dust disk, characterized by a temperature of 800$\pm$300 K and a $L_{dust}/L_{\ast}$ of 5$\pm$3$\times$10$^{-3}$. K2-237 should be a young star, as the isochrone fitting gives a stellar age of 1.0$^{+1.4}_{-0.71}$$\times$10$^{9}$ yr; and the stellar rotation of 5.11$\pm$0.02 days is fast for a F-type star.

An Earth-size planet on an inner orbit could be an alternative physical origin of the TTV. We note that more timing observations, e.g., scheduling CHEOPS \citep{Cheops2021} and SPEARNET \citep{A-thano2022, TransitFit} at certain epochs would be crucial in distinguishing different scenarios.

\section{acknowledgements}

This work made use of numpy \citep{Harris2020}, isochrones\footnote{\url{https://isochrones.readthedocs.io/en}} \citep{Dotter2016}, emcee \citep{emcee}, ARIADNE \citep{Vines2022}, NASA Exoplanet Archive 10.26133/NEA12 table\footnote{\url{https://exoplanetarchive.ipac.caltech.edu/index.html}} \citep{ExoplanetArchive} and PyAstronomy\footnote{\url{https://github.com/sczesla/PyAstronomy}} \citep{pya}. Su-Su Shan and Fan Yang are supported by the Beijing Natural Science Foundation (No. 1244059). Su-Su Shan and Fan Yang acknowledge funding from the CSST MilkyWay and Nearby Galaxies Survey on Dust and Extinction Project CMS-CSST-2021-A09.

\section*{Data Availability}
The data underlying this article are available in the article and in its online supplementary material.

\bibliographystyle{mnras}
\bibliography{ref}
\end{document}